\documentclass[prl,aps,amsfont]{revtex4}
\usepackage{psfig}
\usepackage{graphicx}

\def\textbf#1{{\bf #1}}
\def\>{\rangle}
\def\<{\langle}
\def\beq{\begin{equation}}
\def\eeq{\end{equation}}
\def\be{\begin{equation}}
\def\ee{\end{equation}}
\def\ben{\begin{eqnarray}}
\def\een{\end{eqnarray}}
\def\beqa{\begin{eqnarray}}
\def\eeqa{\end{eqnarray}}
\def\eea{\end{array}}
\def\bea{\begin{array}}
\newcommand{\bei}{\begin{itemize}}
\newcommand{\eei}{\end{itemize}}
\newcommand{\bee}{\begin{enumerate}}
\newcommand{\eee}{\end{enumerate}}
\newcommand{\tr}{{\rm tr}}

\def\tr{{\rm Tr}}

\def\>{\rangle}
\def\<{\langle}

\newtheorem{lemma}{Lemma}

\newtheorem{theorem}{Theorem}
\newtheorem{fact}{Fact}
\newtheorem{proposition}{Proposition}

\begin{document}

\title{On some entropic entanglement parameter}
\author{ Barbara Synak-Radtke$^{(1)}$, \L{}ukasz Pankowski$^{(2)}$, Micha\l{} Horodecki$^{(1)}$ and Ryszard Horodecki $^{(1)}$}

\affiliation{$^{(1)}$Institute of Theoretical Physics and Astrophysics,
University of Gda\'nsk, Poland}
\affiliation{$^{(2)}$Department of Mathematics, Physics and Computer Science,
University of Gda\'nsk, Poland}

\begin{abstract}

In this paper we present the quantity, which is an entanglement parameter. Its origin is very intriguing, because
 its construction is motivated by 
separability criteria based on uncertainty relation. 
We show that this quantity is asymptotically continuous. We also find the lower and upper bounds for it. 
Our entanglement parameter  has the same feature as the coherent information: both can be negative.
 There are also some classes of states for which these quantities coincide with  each other.
\end{abstract}

\maketitle
\section{Introduction}

{\it Quantum entanglement} is a key feature of composite quantum systems playing
 a fundamental role in quantum information theory. In contrast to the pure states
there are still open problems connected with the classification, characterization
and quantification of entanglement of the mixed states (or noisy entanglement).
Thus we need   many different tools to describe the
structure of noisy entanglement. In particular there are various {\it  entanglement  measures}
which tell us how much entanglement is present in a given state in reference 
to the singlet - a maximally entangled state. Entanglement measures must fulfill some natural 
conditions. In particular they should distinguish the nonclassical part of 
correlation between subsystems from the classical one. Therefore entanglement measures must behave
monotonously under LOCC operations. However, there are 
some functions of states which are not LOCC monotones but give us a hint 
on the strength of entanglement and reveal some of its characteristic features.
 We will call them  {\it entanglement parameters}. 

 Now the question arises, what should  we expect from an entanglement parameter? Surely it must be 
 a function, which depends only on a given state. 
Besides, it must be nonpositive for separable states and so its positivity  for a given state indicates
 that the state is entangled. 
Entanglement parameters can increase under LOCC, thus they cannot describe entanglement directly, as it would imply
that entanglement can be increased by classical means.
 However, it is 
plausible
that  normalized entanglement parameters simply underestimate some normalized  entanglement measures,
 where normalized entanglement
parameter (measure) means it is equal to 1 for the singlet state. 
Let us consider in this context an entanglement parameter which is called the {\it coherent information} \cite{szumnielsen,HH94-redun} and 
its  maximum value attainable by LOCC. The maximal value is already an entanglement measure. 
(Notice, that from every entanglement parameter one can obtain
entanglement measure as the maximal value attainable by LOCC.)
One can show, that the maximal value of the coherent information  does not exceed $\log_2 d$, i.e. the value on the singlet 
state.
Thus the coherent information can only underestimate the value of the above entanglement
measure. Let us note that it is important to know the maximal value
of the entanglement measure induced by a given parameter on $\mathcal{C}^2 \otimes \mathcal{C}^2$,
so that we have a reference point.

Let us recall some known entanglement parameters (i.e. functions which fulfill the   above requirements). 
A useful entanglement parameter $M(\rho)$ was introduced 
which characterizes the maximal violation of the Bell-CHSH inequality 
for arbitrary mixed two-qubit states \cite{CHSH}. It depends only on some state parameters 
and contains 
all the information that is needed to decide whether a state violates 
the Bell-CHSH inequality. A closely related entanglement  parameter was defined \cite{bellpar}
as a measure of violation of the Bell-CHSH inequality $B(\rho)= 
\sqrt{\max\{0,M(\rho)-1\}}$,
which for an arbitrary two-qubit pure state equals to measures of entanglement
{\it negativity} \cite{neg1,neg2} and {\it concurrence} \cite{con1,con2}. There is another entanglement 
parameter $N(\rho)$ defined
for an arbitrary two-qubit state connected with possibility of 
teleportation by use of a given state $\rho$ as a quantum channel \cite{parN}. 
In particular it has been shown that every two-qubit state,
which violates the Bell-CHSH inequality offers better fidelity of teleportation 
than  the pure classical channel. Perhaps the most useful 
entanglement parameter in quantum communication is  the  coherent information,
which is closely related to  the conditional entropy.

In this paper we introduce a new entanglement parameter. Its construction is  inspired  by the  separability criteria 
considered in 
\cite{entropic,sepcr}, which  are based on uncertainty relation, where 
the concept of detection of  entanglement is as follows. One or several observables  $ M_i$ are taken and 
the sum of entropies $\sum_i S(M_i)_\rho$ or the sum of variances $\sum_i \delta (M_i)_\rho$. 
For product states a  lower bound is derived
for this sum, which by concavity also holds for separable
states. Thus violation of this lower bound for a given state $\rho $ implies that $\rho $ is entangled.
 Notice that these entropic separability criteria, are different from the other approaches (see for example 
\cite{HH94-redun,HHH96,HH96,cerfadami,aberaja,vol}), because here only 
the probability distribution of the outcomes of a measurement is taken into account, and not the eigenvalues
of the density matrix. 

Our parameter is based on the simplest
possible separability criterion emerging from the above concept,
and is given by the following formula:
\ben
\mathcal{M}(\varrho) =\sup_{\mathcal{P}}(\inf_{\delta \in \mathcal{PS}} H(\delta ,\mathcal{P})
-H(\varrho,\mathcal{P}))
\een
where $\mathcal{PS}$ is the  set of product states, $\mathcal{P}\equiv \{P_i\}$ is a set of projectors representing a
von Neumann measurement and 
$H(\varrho,\mathcal{P})$ is the  Klein entropy, which is equal to the Shannon entropy 
for the probability distribution $p_i=\tr \varrho P_i$ and
\mbox{$H(\{p_i\})=-\sum_i p_i \log_2 p_i$.}
One could here consider also Renyi entropies,
however in this paper we concentrate only on the  Shannon one,
basing on experience \cite{HH94-redun,SW-nature} that quantities built out of the 
Shannon and von Neumann entropies often have operational interpretation.


The quantity $\mathcal{M}(\varrho)$, which we will call the {\it entropic entanglement parameter} (${\cal E}$-parameter), 
has interesting features. 
 For all separable states $\mathcal{M}(\varrho)$ is nonpositive.
It tells us about the "strength" of entanglement, 
because if we want to obtain positive value of $\mathcal{M}(\varrho)$ 
we need a state having enough amount  of entanglement. For instance for part 
of the entangled isotropic states ${\cal E}$-parameter is nonpositive. However, the  ${\cal E}$-parameter
also "feels"  some  other feature 
of entangled states, 
because it relates the greatest difference between entropies 
of a "nearest" separable or product state and a given state after making measurement, so it must
somehow see the structure of the state 
and be connected with  complementarity between eigenbasis of an entangled state and  eigenbasis of a product state.

\section{${\cal E}$-parameter for separable and product states}
In this section we show that our ${\cal E}$-parameter distinguishes the separable states,
 because for them the  ${\cal E}$-parameter  is always negative or equal to zero.

\begin{proposition}
For every separable state $\varrho_{sep}$ the ${\cal E}$-parameter $\mathcal{M}(\varrho_{sep} )$ is less or equal to zero.
\be
\mathcal{M}(\varrho_{sep})\leq 0
\ee
\end{proposition}
{\it Proof.}\\
Let $\mathcal{P}$ be a set of projectors representing a measurement. 
We know that  every separable state can be written as a convex 
mixture 
of product states and the  Klein  entropy for a given measurement is a concave function.
 These facts imply that for every separable state 
$\varrho _{sep}$ we can find  a product state $\varrho_{prod}$ such that 
\be  
H(\varrho_{prod} ,\mathcal{P})\leq  H(\varrho_{sep} ,\mathcal{P})
\ee
So
\be  
\inf_{\delta \in \mathcal{PS}} H(\delta ,\mathcal{P})=\inf_{\delta \in \mathcal{SEP}} H(\delta ,\mathcal{P})
\ee
where $\mathcal{SEP}$ is the  set of separable states. Then for a given measurement $\mathcal{P}$ 
\ben
\inf_{\delta \in \mathcal{PS}} H(\delta ,\mathcal{P})-H(\varrho_{sep},\mathcal{P})=
\inf_{\delta \in \mathcal{SEP}} H(\delta ,\mathcal{P})-H(\varrho_{sep},\mathcal{P})\leq 0 
\een
It turns out  that in particular the  ${\cal E}$-parameter
  is equal to zero for all pure product states, which we are going to
prove below.

\begin{proposition}
For every product pure state $\varrho_{prod}$ the ${\cal E}$-parameter $\mathcal{M}(\varrho_{prod} )$ is equal to zero.
\be
\mathcal{M}(\varrho_{prod})=0
\ee
\end{proposition}
{\it Proof.}\\
From the lemma above we know that
\be
\mathcal{M}(\varrho_{prod})\leq 0
\label{r1}
\ee
Lets consider a measurement $\tilde \mathcal{P}$ which is in eigenbasis of state $\varrho_{prod}$. Then 
\ben
\mathcal{M}(\varrho_{prod}) \geq \inf_{\delta \in \mathcal{PS}} H(\delta ,\tilde \mathcal{P})-H(\varrho_{prod},\tilde \mathcal{P})=
\inf_{\delta \in \mathcal{PS}} H(\delta ,\tilde \mathcal{P})= 0 
\label{r2}
\een
The inequalities (\ref{r1}) and (\ref{r2}) imply that 
\be
\mathcal{M}(\varrho_{prod})=0
\ee

\section{Upper bound for $\mathcal{M}$}

In this section we present the upper bounds
 for our quantity. Some of them are general and some are applicable for a fixed  dimension.

\begin{proposition}
Let $\varrho $ be a state acting on  Hilbert space $\mathcal{C}^d \otimes \mathcal{C}^d$.
Then  $\mathcal{M}(\varrho )$ is bounded from above by 
information content $I(\varrho )$ of state $\varrho $.
\be
\mathcal{M}(\varrho ) \leq I(\varrho )= 2\log_2 d -S(\varrho )
\ee
\end{proposition}
{\it Proof.}\\
We use the fact that for any state $\varrho $ we have 
\be
H(\varrho,\mathcal{P}) \geq S(\varrho )\quad \textrm{and}\quad H(\varrho,\mathcal{P})\leq 2\log_2 d.
\ee
Then
\ben
\mathcal{M}(\varrho) =\sup_{\mathcal{P}}(\inf_{\delta \in \mathcal{PS}} H(\delta ,\mathcal{P})
-H(\varrho,\mathcal{P}))\leq \sup_{\mathcal{P}}\inf_{\delta \in \mathcal{PS}} H(\delta ,\mathcal{P})-S(\varrho )
\leq 2\log_2 d -S(\varrho )
\een
{\it Remark}\\
Notice that this bound is not very good, because for all states it is greater or equal to zero. 
So it does not "see" the difference between a separable state and an entangled one. 
Thus this bound can not  give us any information about the structure of a state. 
However, it is natural and has an operational meaning, because it is equal to information content of state.
Below we present a little better bound for ${\cal E}$-parameter.

\begin{proposition}
Let $\varrho $ be a state acting on  Hilbert space $\mathcal{C}^d \otimes \mathcal{C}^d$.
Then  $\mathcal{M}(\varrho )$ is bounded from above as follows:

\be
\mathcal{M}(\varrho )\leq \log_2d+(1-\frac{1}{d})\log_2(d+ 1)-S(\varrho )
\ee
\end{proposition}
{\it Proof.}\\
We will try to estimate the  quantity 
\be
\sup_{\mathcal{P}}\inf_{\delta \in \mathcal{PS}} H(\delta ,\mathcal{P})
\ee
Let $\mathcal{P}=\{P_i\}$ and $\sigma_\mathcal{P}$ be a  
product state having the greatest projection $p=\tr P_k\sigma_\mathcal{P} $ on 
the the least entangled projectors $P_k$ from $\{P_i\}$ . Then 
\be
\sup_{\mathcal{P}}\inf_{\delta \in \mathcal{PS}} H(\delta ,\mathcal{P})\leq
 \sup_{\mathcal{P}} H( \sigma_\mathcal{P} ,\mathcal{P})
\ee
Notice that $p$ is equal to the square of the greatest Schmidt coefficient  of state 
$\varphi_k$, where  $P_k=|\varphi_k\>\<\varphi_k|$, so
$\frac{1}{d} \leq p \leq 1$. Then
\ben
&&\sup_{\mathcal{P}}H( \sigma_\mathcal{P} ,\mathcal{P})=
 \sup_{\frac{1}{d} \leq p \leq 1}H(p,\frac{1-p}{d^2-1},\frac{1-p}{d^2-1},.....,\frac{1-p}{d^2-1},)=\\
&&H(\frac{1}{d},\frac{1}{d^2+d},\frac{1}{d^2+d},....,\frac{1}{d^2+d})=\log_2d+(1-\frac{1}{d})\log_2(d+ 1)
\een
Using the same arguments as in  the proof of the previous proposition we get
\ben
\mathcal{M}(\varrho) \leq \sup_{\mathcal{P}}(\inf_{\sigma_\mathcal{P} } H(\sigma_\mathcal{P} ,\mathcal{P})
-H(\varrho,\mathcal{P}))\leq \sup_{\mathcal{P}}\inf_{\sigma_\mathcal{P} } H(\sigma_\mathcal{P} ,\mathcal{P})-S(\varrho )
\leq  \log_2d+(1-\frac{1}{d})\log_2(d+ 1)-S(\varrho )
\een
{\it Remark.} Notice that this bound is nonpositive  for the maximally mixed state.
\be
\mathcal{M}(\frac{I}{d^2})\leq -\log_2d+(1-\frac{1}{d})\log_2(d+1)<0
\ee

\begin{proposition}
\label{ogr}
Let $\varrho $ be a state acting on  Hilbert space $\mathcal{C}^2 \otimes \mathcal{C}^2$.
Then  
\be
\mathcal{M}(\varrho ) \leq  1-S(\varrho )
\ee
\end{proposition}
First we show the following fact:
\begin{fact}
\label{f}
In any 2-dimensional subspace  of space $\mathcal{C}^2\otimes\mathcal{C}^2$ we can find a product state.
\end {fact}
{\it Proof.} Let $|\psi_1\>$ and $|\psi_2\>$ be states spanning 2-dimensional subspaces.
\ben 
\psi_1 =\sum_{ij}^2a_{ij}|ij\> \quad \psi_2 =\sum_{ij}^2b_{ij}|ij\>
\een
Let $A=\{a_{ij}\}$ and $B=\{b_{ij}\}$. We show that we can find a product state in 
$span(|\psi_1\>,|\psi_2\>)$ i.e
\ben
\label{rown}
\exists_{\alpha,\beta }\quad \alpha |\psi_1\>+\beta |\psi_2\> =\varphi_{prod} 
\een
 In the space $\mathcal{C}^2\otimes\mathcal{C}^2$ equation (\ref{rown})  is equivalent to the following conditions:
\ben
\exists_{\alpha,\beta }\quad r[\alpha A+\beta B] =1 \Longleftrightarrow 
\exists_{\alpha,\beta }\quad Det[\alpha A+\beta B] =0 
\een
Notice that
\ben
Det[\alpha A+\beta B] =0 \Longleftrightarrow \alpha Det[ A+\gamma  B] =0  \Longleftrightarrow Det[ A+\gamma  B] =0 
\een
where $\gamma=\frac{ \beta}{ \alpha} $.
After elementary  calculations we get the following  equation:
\ben
Det[B]\gamma^2  +c\gamma +d=0
\label{kwad}
\een
Notice that if $B$ is not a matrix representing a product state then $Det[B]\neq 0$. 
So  equation (\ref{kwad}) always has  the solution, because it is the square equation. In the opposite case, if 
$Det[B]=0$ and $|\psi_2\>$ is a product state, we immediately have a product state in $span(|\psi_1\>,|\psi_2\>)$. 

{\it Proof (of Proposition \ref{ogr})}. 
Fact \ref{f} implies that for any measurement $\mathcal{P} \in \mathcal{C}^2\otimes\mathcal{C}^2$, $\mathcal{P}=\{P_i\}_{i=1}^4$
there exists a product state $\delta_{prod}^\mathcal{P} \in span\{P_k,P_l\}$ ($ k,l \in\{1,2,3,4\}$, $k \neq l$). So

\be
\inf_{\delta \in \mathcal{PS}} H(\delta ,\mathcal{P})\leq H(\delta_{prod}^\mathcal{P},\mathcal{P}) \leq 1
\ee
and
\be
\sup_{\mathcal{P}}(\inf_{\delta \in \mathcal{PS}} H(\delta ,\mathcal{P}))=1
\ee
where supremum is achievable in the Bell basis. Then

\ben
\mathcal{M}(\varrho) =\sup_{\mathcal{P}}(\inf_{\delta \in \mathcal{PS}} H(\delta ,\mathcal{P})
-H(\varrho,\mathcal{P}))\leq \sup_{\mathcal{P}}\inf_{\delta \in \mathcal{PS}} H(\delta ,\mathcal{P})-S(\varrho )
=1 -S(\varrho )
\een
{\bf Remark.}
For $\mathcal{C}^3 \otimes \mathcal{C}^3$ we have numerical result saying that
\be
\sup_{\mathcal{P}}(\inf_{\delta \in \mathcal{PS}} H(\delta ,\mathcal{P}))\approx  1.71
\ee
It implies (analogously to above proposition) that
\ben
\mathcal{M}(\varrho) \leq 1.71-S(\varrho )
\een

\section{Lower bounds for ${\cal E}$-parameter}
Let us now pass to the lower bounds, which  we obtained for the value of ${\cal E}$-parameter.

\begin{proposition}
\label{iso}
Let $\varrho^{B}$ be a  state  diagonal in a maximally entangled  basis  on  Hilbert space 
$\mathcal{C}^d \otimes \mathcal{C}^d$.
Then  
\be
\mathcal{M}(\varrho^{B} ) \geq \log_2 d -S(\varrho^{B} )=I_{coh}(\varrho^{B} )
\ee
where $I_{coh}$ is the coherent information equal to a difference between the von Neumann entropy of subsystem and 
 entropy of entire state.
\end{proposition}
{\it Proof.}\\
We know that for any state $\varrho$ if measurement $\mathcal{P}$ is made in the  eigenbasis of the state $\varrho$ 
then the Klein entropy $H(\varrho,\mathcal{P})$ is equal to the  von Neumann entropy of the state. 
A basis consisting of maximally entangled projectors $\mathcal{P}^B \equiv \{P^B_i\}$ (so called Bell basis) is 
eigenbasis of  state $\varrho^{B}$.  
 So
\ben
\mathcal{M}(\varrho^{B}) =\sup_{\mathcal{P}}(\inf_{\delta \in \mathcal{PS}} H(\delta ,\mathcal{P})
-H(\varrho^{B},\mathcal{P}))
\geq \inf_{\delta \in \mathcal{PS}} H(\delta ,\mathcal{P}^B)-S(\varrho^{B} )
\een
Now, we have to show that  $\inf_{\delta \in \mathcal{PS}} H(\delta ,\mathcal{P}^B)=\log_2 d$.
 First notice that $p_i=\tr P^B_i\sigma \in [0,\frac{1}{d}]$ for any product state $\sigma$.
  Notice that above restriction  implies that at least $d$ of probabilities $p_i$
must be nonzero and if exactly $d$ are nonzero than we have $\{p_i\}=\{\frac{1}{d},\frac{1}{d},...,\frac{1}{d}\}$.
Here we can recall the following fact \cite{shannon}, saying that 
any change toward equalization of probabilities $p_1, p_2, \ldots,  p_n$ 
increases $H$ and if $p_1 < p_2$ and we increase $p_1$,
  decreasing $p_2$ by an equal amount so that $p_1$ and $p_2$ are
  more nearly equal, then $H$ increases. This fact implies that
 $H(\{\frac{1}{d}, \frac{1}{d},...,\frac{1}{d}-\delta,\delta \} > H(\{\frac{1}{d},\frac{1}{d},...,\frac{1}{d}\})=\log_2 d$, where $\delta \in (0,\frac{1}{d})$. 
By the induction  we get that the smallest entropy attainable under given restriction is $\log_2 d$.
So we have
\ben
\mathcal{M}(\varrho^{B}) \geq \log_2 d -S(\varrho^{B} )
\een
{\bf Remark 1}. In general, we have a lower bound which is true  for any state:
\ben
\mathcal{M}(\varrho) \geq \inf_{\delta \in \mathcal{PS}} H(\delta ,\mathcal{P}^\varrho )-S(\varrho)
\een
where $\mathcal{P}^\varrho $ represents the measurement made in eigenbasis of state $\varrho $.\\

\begin{lemma}
Let $c$ be the greatest Schmidt coefficient of the state $\varphi $ 
acting on the Hilbert space $\mathcal{C}^d \otimes \mathcal{C}^d$.
If we are able to construct the eigenbasis $\{|\psi_i\>\}$ of $|\psi\>$, 
i.e. $|\psi\> \in \{|\psi_i\>\}$ such that the greatest Schmidt coefficient of 
each $\psi_i$ is less or equal to $c$, then the following inequality holds
\ben
\inf_{\sigma_{prod} }H(\sigma_{prod},P)\geq -k c\log_2c-(1-kc)\log_2(1-kc)=H(c,...,c,1-kc)
\een
where  infimum is taken over all product states $ \sigma_{prod} $, 
 $k$ is equal to $\lfloor \frac{1}{c} \rfloor$ and  $k$ 
is the number of $c$. 
\label{prod}
\end{lemma}
{\it Proof.}\\
Notice that the maximal overlap between the state $\psi_i $ and any product state $\sigma_{prod} $ 
is equal to the square of the 
greatest Schmidt coefficient of $\psi _i$ \cite{overlap}.
 Let $p_i=TrP_i\sigma$, then every probability $p_i$ is bounded by c i.e $p_i\leq c$. 
If we want to have the smallest entropy $H(\{p_i\})$  we must have as many of the probabilities $p_i$ equal to 0 or $c$ 
as possible. 
There may be at most $\lfloor \frac{1}{c} \rfloor$ probabilities $p_i$ equal to $c$.
These conditions are
connected with the concavity of entropy and the lemma that any change of probabilities
towards equalization  increases entropy, which  implies that
\\ 1) $H(c,...,c,1-kc) \leq H(c-\delta,...,c, 1-kc, \delta )$
\\2) $H(c,...,c,1-kc) \leq H(c-\delta,...,c, 1-kc + \delta)$
\\By using the condition 1) and 2) and the induction rule we show that entropy $H(c,...,c,1-kc)$
 is the least achievable for probability distribution 
with all $p_i\leq c$.

\begin{proposition}
\label{czystedo8}
Let $|\varphi\> $ be a pure state  acting on the Hilbert  space $\mathcal{C}^d \otimes \mathcal{C}^d$ where d=2,4,8.
Then  
\ben
\mathcal{M}(\varphi )\geq H(c,...,c,1-kc)
\een
where  $c= max\{a_i^2\}$ for  $|\varphi\> =\sum_{i}a_i|e_i\>\otimes|f_j\>$,
$k$ is the number of probabilities equal to $c$ and  $k=\lfloor \frac{1}{c} \rfloor$.
\end{proposition}
{\it Proof}\\
For the state  $|\varphi\> =\sum_{i}a_i|e_i\>\otimes|f_j\>$
 acting on the Hilbert space $\mathcal{C}^d \otimes \mathcal{C}^d$ 
(where $d=2,4,8$), 
we are able to construct a  measurement  $P^\varphi =\{|\varphi_k\>\<\varphi _k|\}$ 
representing the eigenbasis of the  state 
$|\varphi\> $ such that every Schmidt coefficient of the vector 
$|\varphi_i \>$ is less or equal to the greatest coefficient of $|\varphi \>$. Then by lemma \ref{prod} we have 
\ben
\mathcal{M}(\varphi ) =\sup_{\mathcal{P}}(\inf_{\delta \in \mathcal{PS}} H(\delta ,\mathcal{P})
-H(\varphi ,\mathcal{P}))
\geq \inf_{\delta \in \mathcal{PS}} H(\delta ,P^\varphi)\geq H(c,...,c,1-kc)
\een
Our basis will consist of the vector $|\varphi\>$ and other vectors $|\varphi _k\>$ 
with the same (with regard to absolute value) 
set of Schmidt coefficients in the basis $\{|e_i\>\otimes|f_j\>\}$,
 where we choose such representation of the state $|\varphi\>$ that $a_i \in \mathcal{R}$.
 Then we can construct an  eigenbasis of $|\varphi\>$ consisting of the vectors $|\varphi\>$ 
and other vector $|\varphi _k\>$ with the same (with regard to the absolute value) 
set of Schmidt coefficients in the basis $\{|e_i\>\otimes|f_j\>\}$.
In the case $\mathcal{C}^2 \otimes \mathcal{C}^2$ we can express the basis  $\{|\varphi _k\>\} $ as the following matrix:
$$\pmatrix{
a_{1} & 0& 0& a_{2} \cr
-a_{2} & 0 & 0 & a_{1} \cr
 0 & a_{1}& a_{2}& 0 \cr
0 &  -a_{2} &a _{1}& 0 \cr
}$$
The analogous matrix representing the eigenbasis of $|\varphi\> $ for the Hilbert space  
$\mathcal{C}^4 \otimes \mathcal{C}^4$ is of the form:

$$\pmatrix{
a_{1}& 0& 0& 0& 0 & a_{2} & 0 & 0& 0& 0& a_{3}& 0& 0 & 0 & 0& a_{4} \cr
-a_{2}& 0 &  0 &  0 &  0 &  a_{1} & 0 &  0 &  0 &  0 &  a_{4} & 0 &  0 &  0 &  0 &  -a_{3} \cr
a_{3} &0 &  0 &  0 &  0 & a_{4} &0 &  0 &  0 &  0 & -a_{1}&0 &  0 &  0 &  0 & -a_{2} \cr
a_{4} &0 &  0 &  0 &  0 & -a_{3}&0 &  0 &  0 &  0 & a_{2} & 0 &  0 &  0 &  0 &-a_{1} \cr
 0 &a_{1} &  0 &  0 &  a_{2} &0 &   0 &  0 &  0 &  0 &  0 & a_{3} & 0 &  0 &  0 &  a_{4}  \cr
 0 & -a_{2}& 0 &  0 &  a_{1} &0 &   0 &  0 &  0 &  0 &   0 & a_{4}  &  0 &  0 &  0 &  -a_{3} \cr
0 &a_{3} &  0 &  0 &a_{4}&  0  & 0 &  0 &  0 &  0 & 0 &-a_{1}&  0 &  0 &  0 & -a_{2} \cr
0 & a_{4} & 0 &  0 &  -a_{3}&0 & 0 &  0 &  0 &  0 &  0 &a_{2} &  0 &  0 &  0 &-a_{1} \cr
 0 &  0 &a_{1} &  0 &  0 &   0 &  0 &a_{2} &  a_{3} &0 &  0 &  0 &  0 &  a_{4} &  0 &  0  \cr
 0 &  0 &-a_{2}&  0 &  0 &   0 &  0 &a_{1} &   a_{4} &0 &  0 &  0 &  0 &  -a_{3}&  0 &  0  \cr
0 &  0 &a_{3} &  0 &  0 & 0 &  0 &a_{4} & -a_{1}& 0 &  0 & 0 &  0 & -a_{2}&  0 &  0  \cr
0 &  0 &a_{4} &  0 &  0 & 0 &  0 & -a_{3}&a_{2} & 0 &  0 &  0 &  0 &-a_{1} &  0 &  0 \cr
 0 &  0 &  0 &a_{1} &  0 &   0 & a_{2} & 0 &  0 &  a_{3} &0 &  0 &  a_{4}&  0 &  0 &  0   \cr
 0 &  0 &  0 &-a_{2}&  0 &   0 &a_{1} &  0 &  0 &  a_{4} &0 &   0 &  -a_{3}&  0 &  0 &  0  \cr
0 &  0 &  0 &a_{3} &  0 & 0 &a_{4} &  0 &  0 & -a_{1}& 0 & 0 & -a_{2}&  0 &  0 &  0  \cr
0 &  0 &  0 & a_{4} & 0 & 0 &-a_{3}&  0 &  0 & a_{2} & 0 &  0 &-a_{1}&  0 &  0 &  0  \cr
}$$
For $\mathcal{C}^8 \otimes \mathcal{C}^8$ this basis will consist of 8 groups of vectors each spanning an 
orthogonal subspace. Values of  non-zero coefficients of the vectors of every group rewritten 
in basis $\{|e_i\>\otimes|f_j\>\}$ 
 represents the following matrix: 

$$\pmatrix{
a_1& a_2& a_3& a_4& a_5& a_6& a_7& a_8  \cr
a_2& -a_1& -a_4& a_3& -a_6& a_5& -a_8& a_7  \cr
a_3& a_4& -a_1& -a_2& a_7& -a_8& -a_5& a_6  \cr
-a_4& a_3& -a_2& a_1& a_8& a_7& -a_6& -a_5  \cr
-a_5& -a_6& a_7& -a_8& a_1& a_2& -a_3&a_4  \cr
-a_6& a_5& -a_8& -a_7& -a_2& a_1& a_4& a_3  \cr
a_7& a_8& a_5& -a_6& -a_3& a_4& -a_1& -a_2  \cr
-a_8& a_7&  a_6& a_5& -a_4& -a_3& -a_2& a_1  \cr
}$$
{\bf Remark 1.}
In particular as a consequence of proposition \ref{czystedo8} for any pure state $|\varphi\> $ 
acting on the Hilbert  space $\mathcal{C}^2 \otimes \mathcal{C}^2$ 
we get the following lower bound for ${\cal E}$-parameter.
\be
\mathcal{M}(\varphi ) \geq E(\varphi )=S_R(\varphi )
\ee
where $E $ is entanglement measure for pure bipartite state, which is  equal to $S_R(\varphi )$ -
the  von Neumann entropy of reduced density matrix $\tr_B|\varphi \>\<\varphi|= \tr_A|\varphi \>\<\varphi|$.
 \\{\bf Remark 2.} Notice that this lower bound is not greater than entropy of subsystem of state $|\varphi\>$, 
so is not greater than
 coherent information.




\section{Results for some families of states }
Evaluating ${\cal E}$-parameter is very difficult, because its  definition is a kind of so called "minmax".
But there are some classes of states, in particular  states with high symmetry, for which  we are able to find the exact  value of $\mathcal{M}$.

For states diagonal in maximally entangled basis consisting of the vectors  
acting on Hilbert space $\mathcal{C}^2 \otimes \mathcal{C}^2$, we know the value of ${\cal E}$-parameter:
\be 
\mathcal{M}(\varrho ^{B} )= 1 - S(\varrho^{B} )
\ee
which in particular gives a result for  maximally entangled state $\psi_+^2 $
\be
\mathcal{M}(\psi_+^2 )= 1
\ee
This result follows from combining the lemma \ref{iso}
 with  the lemma \ref{ogr}.
\\For  maximally entangled states in higher dimension than d=2 
 we only have a lower bound
\be 
\mathcal{M}(\psi_+^d )\geq \log_2 d
\ee
For $d= 3$ we have numerical result 
\be 
\mathcal{M}(\psi_+^3 )\approx 1.663 > \log_2 3
\ee
 We know the value of ${\cal E}$-parameter for $\mathcal{C}^2 \otimes \mathcal{C}^2$ isotropic states $\varrho_{iso}$,
 which belong to  a subset of Bell diagonal states:

\be
\mathcal{M}(\varrho_{iso}) =  1- S(\varrho^{iso} )
\ee
where the isotropic states are of the form
 \ben
\varrho_{iso}= p P_{+} + \frac{1-p}{d^2}I \quad  \lambda \in [0,1]
\een
and  $P_{+}$ is maximally entangled state and $I$ is the identity matrix. 

\begin{figure}[h]
\begin{center}
\psfig{file=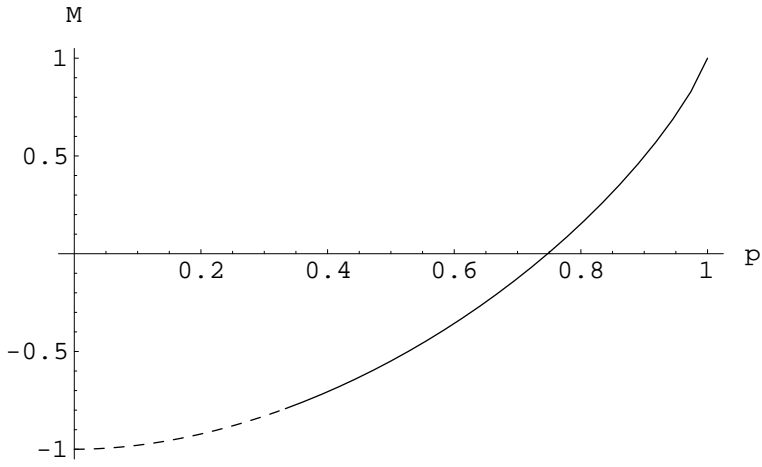}
\end{center}
\caption{The dashed line represents the value  of ${\cal E}$-parameter for the separable isotropic states 
 and the solid one the value   
for entangled isotropic states. \label{rys}}. 
\end{figure}

Figure \ref{rys} shows the value  of ${\cal E}$-parameter for the  isotropic states. 
We can see that for the separable isotropic states
$\mathcal{M}<0$, but what interesting for a part of the entangled isotropic states $\mathcal{M}$ is negative. 
So it means that a given state  must have "enough" entanglement to  have positive value of our entanglement parameter.


{\bf Negative result}. We  have suspected that for a pure state $\varphi$, we get $ \mathcal{M}(\varphi )=S_A(\varphi )$. 
But numerical calculations show that there exist such states (and for a set of randomly chosen states it turns out
to be the majority of them) 
for which
\be 
 \mathcal{M}(\varphi )\neq S_A(\varphi )
\ee

\section{Asymptotic continuity of $\mathcal{M}$} 

Our entanglement parameter has a feature, which is especially useful in the regime of many copies, i.e.
asymptotic continuity. 

\begin{theorem}
\label{teo}
For any state $\varrho $ the quantity $\mathcal{M}(\varrho)$  is asymptotically continuous, which refers the condition
\ben
\forall_{\varrho_1,\varrho_2}  | \mathcal{M}(\varrho_1)-\mathcal{M}(\varrho_2) |\leq K \varepsilon \log_2d+C
\label{asc}
\een
where C is constant and $\varepsilon =||\varrho_1-\varrho_2||$.
\end{theorem}
{\it Proof.}
We show  that $\mathcal{M}$ is "robust under admixture" i.e
\ben
|(  \mathcal{M}( \varrho )-\mathcal{M}((1-\varepsilon )\varrho +\varepsilon \sigma )|\leq 4\varepsilon \log_2d+H(\varepsilon) 
\label{rua}
\een
This feature is equivalent to asymptotic continuity, which  is proven in paper \cite{ascont}.\\
Before we pass to the proof of the theorem we need to introduce the following lemma:
\begin{lemma}
\label{ll}
Let $\mathcal{P}$ be a given measurement. Then
\be
|\mathcal{M}(\varrho,\mathcal{P})-\mathcal{M}(((1-\varepsilon )\varrho +\varepsilon\sigma ) ,\mathcal{P})|\leq 4\varepsilon \log_2d+H(\varepsilon) 
\ee
where
\be
\mathcal{M}(\varrho ,\mathcal{P})=\inf_{\delta \in \mathcal{PS}} H(\delta ,\mathcal{P})
-H(\varrho,\mathcal{P})
\ee
\end{lemma}
{\it Proof.}
\ben \nonumber
&&|\mathcal{M}(\varrho,\mathcal{P})-\mathcal{M}(((1-\varepsilon )\varrho +\varepsilon\sigma ),\mathcal{P})|=
|\inf_{\delta \in \mathcal{PS}} H(\delta ,\mathcal{P})-
H(\varrho,\mathcal{P})-\inf_{\delta \in \mathcal{PS}} H(\delta ,\mathcal{P})+H(((1-\varepsilon )\varrho +\varepsilon\sigma) ,\mathcal{P})|\\ \nonumber
&&=|H(((1-\varepsilon )\varrho +\varepsilon\sigma)  ,\mathcal{P})-H(\varrho ,\mathcal{P})|
=|H(((1-\varepsilon )\varrho +\varepsilon\sigma),\mathcal{P})-(1-\varepsilon )H(\varrho ,\mathcal{P})-\varepsilon H(\sigma  ,\mathcal{P}) 
-\varepsilon H(\varrho ,\mathcal{P})+\varepsilon H(\sigma  ,\mathcal{P})|\\
&&\leq |H(((1-\varepsilon )\varrho +\varepsilon\sigma),\mathcal{P})-(1-\varepsilon )H(\varrho ,\mathcal{P})-\varepsilon H(\sigma  ,\mathcal{P})|+ 
\varepsilon |H(\varrho ,\mathcal{P})|+\varepsilon |H(\sigma  ,\mathcal{P})|\leq H(\varepsilon)+4\varepsilon \log_2d
\een
We use here the facts that $|H(\varrho )|\leq 2 \log_2d$ and 

\be
\sum_k p_kH(\varrho _k,\mathcal{P})\leq H(\sum_k p_k\varrho _k,\mathcal{P})\leq \sum_k p_kH(\varrho _k,\mathcal{P})+H(\{p_k\} )
\ee
which implies 
\be
|H(\sum_k p_k\varrho _k,\mathcal{P})-\sum_k p_kH(\varrho _k,\mathcal{P})|\leq H(\{p_k\} )
\ee
{\it Proof of theorem \ref{teo}.}
\\ 
From lemma \ref{ll} we have   that 
\be
\mathcal{M}(\varrho,\mathcal{P})-\mathcal{M}(((1-\varepsilon )\varrho +\varepsilon\sigma  ) ,\mathcal{P})\leq 4\varepsilon \log_2d+H(\varepsilon)
\ee
then 
\be
\mathcal{M}(\varrho,\mathcal{P})\leq \mathcal{M}(((1-\varepsilon )\varrho +\varepsilon\sigma  ) ,\mathcal{P})+4\varepsilon \log_2d+H(\varepsilon)
\ee
Let $\mathcal{P}_\varrho$ be a measurement achieving $\mathcal{M}(\varrho)$. Then

\ben
\mathcal{M}(\varrho,\mathcal{P}_\varrho )=\mathcal{M}(\varrho)\leq \mathcal{M}(((1-\varepsilon )\varrho +\varepsilon\sigma  ) ,\mathcal{P})+4\varepsilon \log_2d+H(\varepsilon) 
\leq \mathcal{M}((1-\varepsilon )\varrho +\varepsilon\sigma  ) +4\varepsilon \log_2d+H(\varepsilon) 
\een
so
\ben
\mathcal{M}(\varrho)-\mathcal{M}((1-\varepsilon )\varrho +\varepsilon\sigma  )\leq 4\varepsilon \log_2d+H(\varepsilon) 
\label{1}
\een
Analogously we can show that 
\ben
\mathcal{M}(\varrho)-\mathcal{M}((1-\varepsilon )\varrho +\varepsilon\sigma  ) \geq -(4\varepsilon \log_2d+H(\varepsilon) )
\label{2}
\een
Inequalities (\ref{1}) and (\ref{2}) together give us inequality (\ref{rua}), 
which is equivalent to the one from theorem \ref{teo}, 
which  ends the proof.

\section{Conclusion} 
We have introduced a new quantity - an entropic entanglement parameter (${\cal E}$-parameter), 
which has the same feature as coherent information: both can be negative. More precisely, 
we have shown, 
that for all separable states it is always nonpositive  and indeed happens to be negative, which we have shown
for maximally mixed state.
Moreover, we have proved, 
that ${\cal E}$-parameter is asymptotically continuous and we have obtained  upper and lower bounds
 for some classes of  states. The ${\cal E}$-parameter is rather difficult to deal with, which is caused by its definition being a kind of so called "minmax".
Note that, the parameter is not LOCC monotone. It follows from two facts: one that we can pass from 
any separable state to other one using  LOCC operations and second that ${\cal E}$-parameter 
has not the same value for all separable states.  

There are still many
 open questions. We would like to know how  the value of $\mathcal{M}$ can change
 if in  definition we take supremum over POVMs instead of only von Neumann measurements, in particular, 
whether is it possible to obtain infinity. There is also an interesting question 
whether the ${\cal E}$-parameter is, in general, bounded from below by coherent information,
which  we have proven for some  classes of states. 

Finally, we believe that entropic entanglement parameter may reveal some new feature of entanglement
 as it feels the structure of the state and is connected with complementarity
 between eigenbasis of an entangled state and a product one.

{\bf Acknowledgments.} 
We would like to thank Karol Horodecki for helpful discussion.
This work is supported by Polish Ministry of Scientific Research and Information Technology under 
the (solicited) grant no. PBZ-MIN-008/P03/2003, EU grants RESQ (IST-2001-37559), 
QUPRODIS (IST-2001-38877) and EC IP SCALA.

\bibliography{refbasia} 
\end{document}